\documentclass[12pt,preprint]{aastex}

\usepackage{graphicx}

\shorttitle{The \emph{Kepler} Pixel Response Function}
\shortauthors{Bryson et al.}

\begin{document}


\title{The \emph{Kepler} Pixel Response Function}


\author{Stephen T. Bryson\altaffilmark{1},
Peter Tenenbaum\altaffilmark{2},
Jon M. Jenkins\altaffilmark{2},
Hema Chandrasekaran\altaffilmark{2},
Todd Klaus\altaffilmark{3},
Douglas A. Caldwell\altaffilmark{2},
Ronald L. Gilliland\altaffilmark{4},
Michael R. Haas\altaffilmark{1},
Jessie L. Dotson\altaffilmark{1},
David G. Koch\altaffilmark{1},
William J. Borucki\altaffilmark{1}
}


\altaffiltext{1}{NASA Ames Research Center, Moffett Field, CA 94035}
\altaffiltext{2}{SETI Institute, Mountain View, CA 94043}
\altaffiltext{3}{Orbital Sciences Inc., Moffett Field, CA 94035}
\altaffiltext{4}{Space Sciences Telescope Institute, Baltimore, MD 21218}


\begin{abstract}
\emph{Kepler} seeks to detect sequences of transits of Earth-size exoplanets orbiting Solar-like stars. 
Such transit signals are on the order of 100 ppm. The high photometric precision demanded by 
\emph{Kepler} requires detailed knowledge of how the \emph{Kepler} pixels respond to starlight during a 
nominal observation. This information is provided by the \emph{Kepler} pixel response 
function (PRF), defined as the composite of \emph{Kepler's} optical point spread function, integrated 
spacecraft pointing jitter during a nominal cadence and other systematic 
effects. To provide sub-pixel resolution, the PRF is represented as a piecewise-continuous 
polynomial on a sub-pixel mesh. This continuous representation allows the prediction of a star's 
flux value on any pixel given the star's pixel position. The advantages and difficulties of this 
polynomial representation are discussed, including characterization of spatial variation in the 
PRF and the smoothing of discontinuities between sub-pixel polynomial patches. On-orbit 
super-resolution measurements of the PRF across the \emph{Kepler} field of view are described. 
Two uses 
of the PRF are presented: the selection of pixels for each star that maximizes the 
photometric signal to noise ratio for that star, and PRF-fitted centroids which provide robust and 
accurate stellar positions on the CCD, primarily used for attitude and plate scale tracking. 
Good knowledge of the PRF has been a 
critical component for the successful collection of high-precision photometry by \emph{Kepler}. 
\end{abstract}


\keywords{planetary systems --- techniques: photometric}

%%%%%%%%%%%%%%%%%%%%%%%%%%%%%%%%%%%%
%%%%%%%%%%%%%%%%%%%%%%%%%%%%%%%%%%%%
%%%%%%%%%%%%%%%%%%%%%%%%%%%%%%%%%%%%
%%%%%%%%%%%%%%%%%%%%%%%%%%%%%%%%%%%%
%%%%%%%%%%%%%%%%%%%%%%%%%%%%%%%%%%%%
%%%%%%%%%%%%%%%%%%%%%%%%%%%%%%%%%%%%
\section{Introduction} \label{section:intro}
The high photometric precision demanded by 
\emph{Kepler} \citep{borucki10, koch10} requires detailed knowledge of how pixels respond to starlight during a 
nominal observational interval. This is provided by the \emph{Kepler} pixel response 
function (PRF), a super-resolution representation of the interaction of starlight
with pixels that includes modulation by pointing jitter and other systematic effects during 
an observation (this usage of PRF differs from that in \citep{Lauer}). 
The PRF provides a continuous representation that allows the 
prediction of a star's flux value on any pixel given the star's pixel position: these pixel values are 
samples of the PRF at the star's sub-pixel position.

The \emph{Kepler} focal plane \citep{argabright08,caldwell10} consists of 42 $1024 \times 2200$ CCDs, 
each of which is divided into two $1024 \times 1100$ output channels.  A PRF model is defined on 
each of these 84 output channels, accounting for CCD tip/tilt
and other channel-level variations.  Due to variation in the 
optical PSF within an output channel, the PRF model is pixel-position 
dependent  
(\S\ref{section:prfRepresentation}).  Data is collected at a 29.4 minute cadence;
we refer to this sampling interval as a ``long cadence''.  Bandwidth limitations prevent the storage
and downlinking of all $95 \times 10^6$ pixels at each long cadence:
at most $5.4 \times 10^6$ pixels are downlinked per long cadence.  

The PRF currently contributes to \emph{Kepler's} precision by supporting the selection
of optimal pixels for aperture photometry (\S\ref{section:optimalPixels}), and providing
high-precision PRF-fitted centroids (\S\ref{section:stellarCentroids}) for
attitude determination, crucial for removing the effects of pointing jitter from photometry.
These are sufficient to attain \emph{Kepler's} current high precision \citep{jenkins10a}. 
The future use of differential image analysis will use the PRF to perform target and pixel
level sensitivity corrections, including local image motion, further increasing \emph{Kepler's} precision.

A star's position on the \emph{Kepler} focal plane 
is computed with the software package \emph{raDec2Pix}, developed by the 
\emph{Kepler} Science Operations Center (SOC).  This package uses the 
\emph{Kepler} Focal Plane Geometry (FPG) model (\S\ref{section:fpg}), 
as well as Kepler optical, nominal spacecraft attitude, and velocity aberration models.

%%%%%%%%%%%%%%%%%%%%%%%%%%%%%%%%%%%%
%%%%%%%%%%%%%%%%%%%%%%%%%%%%%%%%%%%%
%%%%%%%%%%%%%%%%%%%%%%%%%%%%%%%%%%%%
%%%%%%%%%%%%%%%%%%%%%%%%%%%%%%%%%%%%
%%%%%%%%%%%%%%%%%%%%%%%%%%%%%%%%%%%%
%%%%%%%%%%%%%%%%%%%%%%%%%%%%%%%%%%%%
\section{The Pixel Response Function} \label{section:prf}

\begin{figure}
\epsscale{1}
\plotone{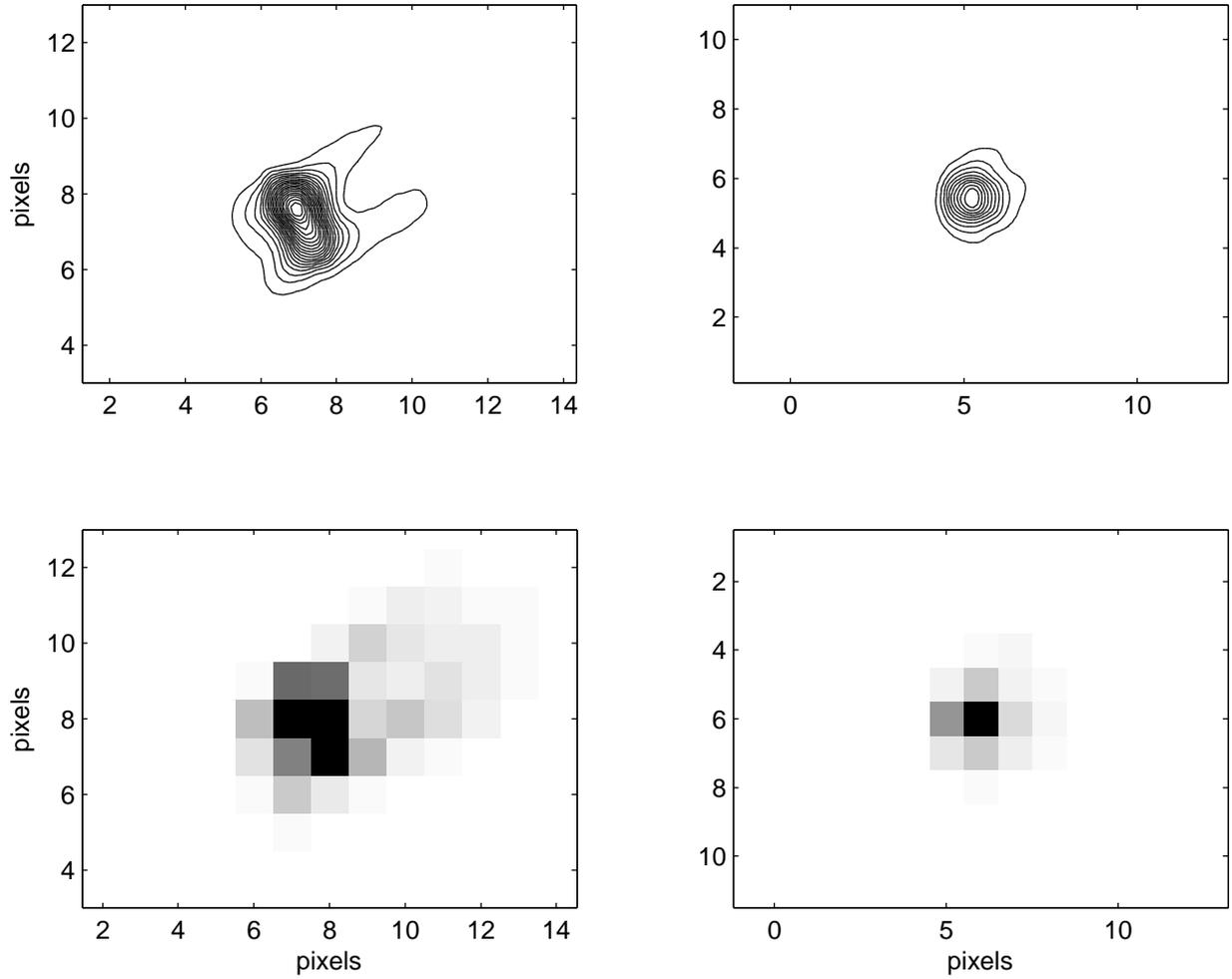}
\caption{Two example PRFs.  Left column: a PRF near the edge of the focal plane.  
Right column: a PRF near the focal plane center.  Top row: the PRF model contours.
Bottom row: the PRF converted into pixel values for a star centered on a pixel.}\label{fig1}
\end{figure}

The \emph{Kepler} PRF on an output channel is represented as a continuous piece-wise polynomial
function of sub-pixel position on each pixel, providing a 
super-resolution image of the light of a star falling on the pixels.
Pixel values are determined by evaluating the PRF given the pixel and sub-pixel
position of a star: the pixel position determines where the PRF is placed in the output
channel's pixel array, while the sub-pixel position determines the pixel values.  
Figure~\ref{fig1} shows two example PRFs and corresponding pixel values.

Computation of the PRF from stellar images requires knowing where those images fall on the
focal plane.  Each star's coordinates are
provided by the Kepler Input Catalog\footnote{http://archive.stsci.edu/kepler/kepler\_fov/search.php} (KIC),  and 
the physical location of the CCDs relative to each other provided by the \emph{focal plane geometry} (FPG) model
(\S\ref{section:fpg}).  Determination of the FPG model, however, requires PRF-fitted stellar centroids 
(\S\ref{section:stellarCentroids}),
so FPG and PRF are mutually dependent.  Therefore we determine both FPG and PRF 
in an iterative process (\S\ref{section:prfFpgIteration}).  

%%%%%%%%%%%%%%%%%%%%%%%%%%%%%%%%%%%%
%%%%%%%%%%%%%%%%%%%%%%%%%%%%%%%%%%%%
\subsection{Representation of the PRF} \label{section:prfRepresentation}

A \emph{Kepler} PRF is defined on a grid of $n \times n$ pixels called the \emph{PRF pixel array}, where 
$n$ is large enough to capture the full PRF.  Each pixel in the PRF pixel array is
sub-divided into $m \times m$ sub-pixel regions.  Each sub-pixel region is
assigned a two-dimensional polynomial $\mathrm{PRF}_{i,j,s,t}\left(x,y \right)$,
where $\left(i,j\right)$ are the pixels indices in the $n \times n$ PRF pixel array and 
$\left(s,t\right)$ are the sub-pixel indices in the $m \times m$ sub-pixel 
grid on pixel $\left(i,j\right)$.  Therefore there are $n^2 \times m^2$ two-dimensional
polynomials associated with a PRF on an output channel.  For most channels $n=11$ is
sufficient to capture the PRF, while for some channels $n=15$ is required.  We find $m=6$
provides sufficient sub-pixel detail.  The order of each
of these polynomials is determined by a maximal information criterion (\S\ref{section:prfComputation}).  

Each pixel of the PRF pixel array gives the flux on that pixel from a star whose location is in the central pixel 
of the PRF pixel array.  Thus the central pixel represents the peak of the star's PRF while pixels towards 
the edge represent the flux in the wings of that star's PRF.  The values of these fluxes depend on the sub-pixel 
position of the star.

PRF polynomials are determined independently of each other, so they will not match
at sub-pixel boundaries.  We handle the resulting discontinuities using two strategies:
(1) the discontinuities are minimized by fitting the PRF using data that extends into 
adjacent sub-pixel regions as described in \ref{section:prfComputation} and (2) 
the remaining discontinuities are smoothed when evaluating the PRF as described in 
\S\ref{section:prfToPixels}.  

A position-dependent PRF is implemented by measuring five PRFs on each
output channel: one at each corner and one at the center.  
Linear interpolation on the resulting triangles as 
described in \S\ref{section:prfToPixels}
provides an interpolated PRF at any point in the output channel.  This model neglects pixel-level variations, which 
are treated during calibration \citep{jenkins10b}.

%%%%%%%%%%%%%%%%%%%%%%%%%%%%%%%%%%%%
%%%%%%%%%%%%%%%%%%%%%%%%%%%%%%%%%%%%
\subsection{Focal Plane Geometry} \label{section:fpg}

The focal plane geometry (FPG) model represents each of the 42 CCDs on the \emph{Kepler} focal plane via the center 
position, rotation angle, and plate scale of each CCD.  Prior to launch the CCD positions were known with an 
accuracy of $\pm 3$ pixels; the FPG model provides the required accuracy of $\pm 0.1$ pixels.  The input to the FPG model computation is a two dimensional polynomial which maps right ascension 
and declination to pixel location.  These {\it motion polynomials} are derived via a fit to the observed 
pixel positions of known stars via PRF centroiding (\S\ref{section:stellarCentroids}), and 
serve as a 
smoothing filter, reducing the impact of observational errors of individual stars.

FPG computation determines the CCD locations of a set of sky coordinates in two ways:  via the {\it 
raDec2Pix} library, which uses the FPG model; and via the motion polynomials, which do not use the FPG model.  
The FPG model parameters are adjusted via a Levenberg-Marquardt non-linear least squares algorithm until the 
differences between the model and observation are minimized.  The FPG model computation also provides the 
spacecraft attitude.

%%%%%%%%%%%%%%%%%%%%%%%%%%%%%%%%%%%%
%%%%%%%%%%%%%%%%%%%%%%%%%%%%%%%%%%%%
\subsection{From PRF to Pixel Values} \label{section:prfToPixels}
The PRF provides expected pixel values for a given star with a 
given magnitude and sky location.  The output channel and pixel
position of the star is determined via the \emph{raDec2Pix} library.  This
pixel position determines which triangle contains the star's central pixel, which 
determines which three (of five) PRFs are used to compute the star's final image. 

Pixel values for the star are determined from each of the three PRFs
by evaluating the PRF polynomials at the sub-pixel position of the star.  On 
each pixel in the PRF pixel array the sub-pixel region containing the star's sub-pixel
coordinates is chosen.  The polynomial in that sub-pixel region is evaluated at
the star's sub-pixel location, providing the relative flux for that pixel.  This
results in an array of $n \times n$ pixels reflecting the relative flux of the star based on that PRF.  

When the sub-pixel location of a star is close to the boundary between
sub-pixel patches, the following method
smoothes out inter-patch discontinuities.  The pixel value is evaluated using all nearby 
polynomial patches (two patches near an edge, four patches near a corner).   
Adjacent polynomial patches can be evaluated at these positions because patches
are defined using data that extend into adjacent patches (\S\ref{section:prfComputation}).
For simplicity we consider the case of an edge, where we have two values $v_1$ and $v_2$ from the 
two polynomial patches adjacent at that edge.  These values are smoothly interpolated using the distance
$z$ of the star's sub-pixel position from the edge with the formula 
$v = w\left( z \right) v_1 + \left( 1-w\left( z \right) \right) v_2$.
Here the weight $w\left( z \right)$ is given, assuming $z$ is normalized so that it ranges from 0 on one
side of the edge to 1 on the other side of the edge, by \citep{warner83}
\begin{equation}
w\left( z \right) = \frac{f\left( z \right)}{f\left( z \right) + f \left(1 - z \right)}, \qquad
f\left( z \right) = \left\{
				\begin{array}{ll}
					\exp\left( -1/z^a \right) & z > 0 \\
					0 & z \leq 0.
				\end{array}
				\right.
\end{equation}
The scale factor $a$ determines the steepness of the exponential curve.
$w\left( z \right)$ has the property that it is equal to 0 for $z \leq 0$, equal to 1 for $z \geq 1$, 
and equal to 0.5 for $z = 0.5$.  Further, $w\left( z \right)$ has continuous derivatives of all order, even at $z = 0$
and $z = 1$.  This approach smoothly eliminates the discontinuity at the boundary between the polynomial patches.

\begin{figure}
\epsscale{1}
\plotfiddle{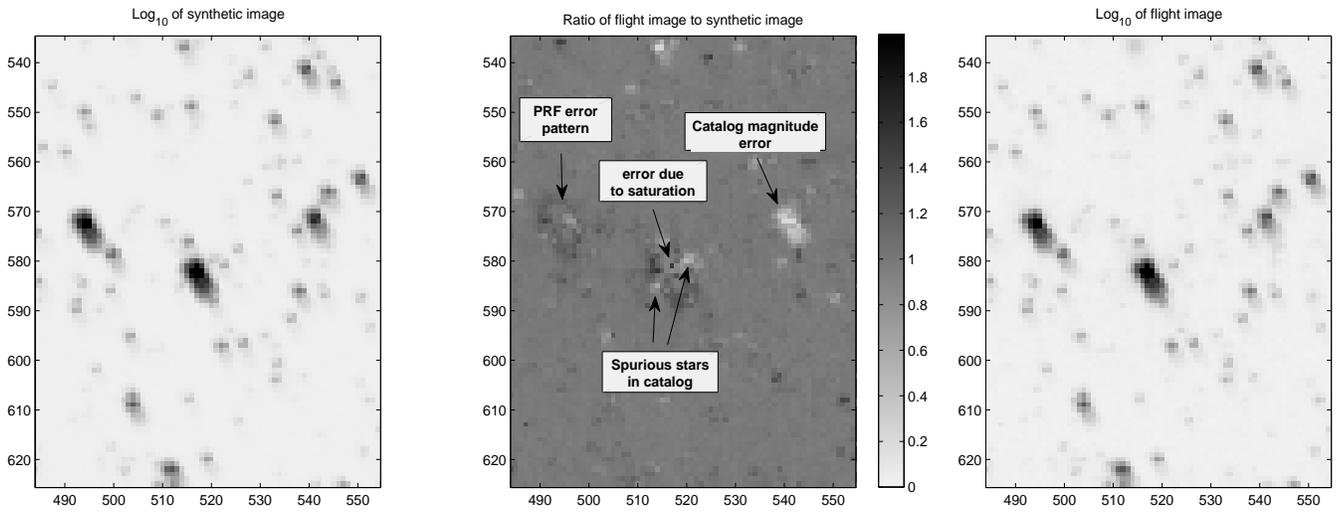}{0in}{0.}{640}{210}{-85}{0}
\caption{Left: Synthetic image created using a PRF and
the Kepler input catalog.  Center: Ratio of the the synthetic image and the flight image, indicating
various examples of error sources.  
Right: Flight image for the same region of sky.}\label{fig2}
\end{figure}

The above process is repeated for all three PRFs defined on the triangle containing
the star's central pixel, giving three $n \times n$ pixel arrays, one at the location 
of each PRF.  These pixel arrays are 
linearly interpolated in the plane of the triangle, giving the final relative pixel values for this star.
These relative pixel values are multiplied by the total flux of the star in the \emph{Kepler} bandpass via the 
\emph{Kepler} magnitude given by the Kepler input catalog.  The final pixel
array is then placed in the output channel pixel array in the location corresponding to the
pixel position of the star.  Repeating this process for every star on an output channel
results in a synthetic image that should match the actual image observed in
flight, assuming the KIC is accurate (Figure~\ref{fig2}).  

%%%%%%%%%%%%%%%%%%%%%%%%%%%%%%%%%%%%
%%%%%%%%%%%%%%%%%%%%%%%%%%%%%%%%%%%%
%%%%%%%%%%%%%%%%%%%%%%%%%%%%%%%%%%%%
%%%%%%%%%%%%%%%%%%%%%%%%%%%%%%%%%%%%
%%%%%%%%%%%%%%%%%%%%%%%%%%%%%%%%%%%%
%%%%%%%%%%%%%%%%%%%%%%%%%%%%%%%%%%%%
\section{Measurement of the PRF} \label{section:prfMeasurement}

The PRF was measured during the commissioning phase of the \emph{Kepler} mission,
after the photometer was brought into final focus.  The PRF measurement
strategy was to collect observations of bright, uncrowded, unsaturated
stars, which are used to fit the PRF polynomial patches.  

%%%%%%%%%%%%%%%%%%%%%%%%%%%%%%%%%%%%
%%%%%%%%%%%%%%%%%%%%%%%%%%%%%%%%%%%%
\subsection{In-flight Observations} \label{section:prfObservations}

\begin{figure}
\epsscale{1}
\plotone{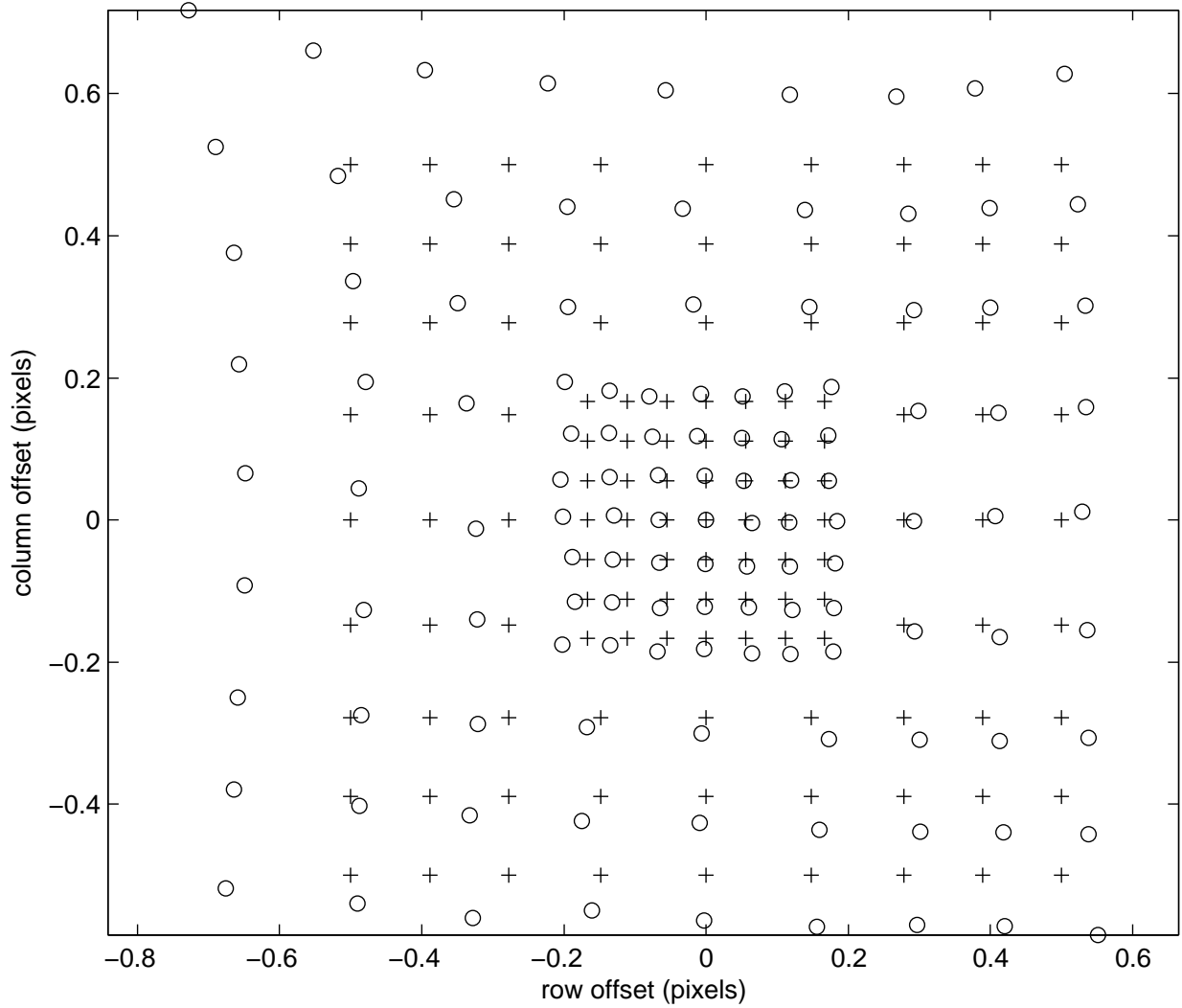}
\caption{Commanded ($+$) and achieved ($\circ$) dither pattern used to measure the PRF.  The 
discrepancy is due to the early state of calibration of the fine guidance trackers during commissioning.}\label{fig3}
\end{figure}

The PRF measurement used 19,189 stars in the \emph{Kepler} field with magnitudes between 12 and 13 that
did not have significant ($> 30\%$) flux from other stars in a $21 \times 21$ pixel aperture
centered on the star.  These stars were observed for 242 cadences of 14.7 minute 
duration, half a long cadence, to reduce the time required
to perform the observations.  Data for the PRF measurement were obtained from each 
odd cadence, while the spacecraft slewed to a new dither location during the even 
cadences.  These 121 PRF data cadences visited the achieved locations shown in 
Figure~\ref{fig3}, which were used in the PRF computation.
The order of visitation of this pattern
was randomized to minimize the impact of time-varying systematics. 
The pixel values for each star
are converted to relative pixel values by normalizing by the total flux from that star.
The pixels from each star measurement of the 121 data cadences are treated as independent data points,
resulting in 2,321,869 observations distributed on 84 output channels.  The distribution 
was uneven, with some channels having many more stars than others.  

%%%%%%%%%%%%%%%%%%%%%%%%%%%%%%%%%%%%
%%%%%%%%%%%%%%%%%%%%%%%%%%%%%%%%%%%%
\subsection{The PRF Computation} \label{section:prfComputation}

Two types of PRF are computed for each output channel: a single PRF using all targets,
and a set of five PRFs to capture intra-channel PRF variation as described in
\S\ref{section:prfToPixels}.  When computing the five-PRF set, the data
for each of the five PRFs are selected from a region near where
that PRF is defined.  For each corner PRF the data is selected from a 
square extending from that corner into the input channel.  For the 
central PRF the data is selected from a square centered on the output channel.
The size of each square is set to include a specified minimum number of
targets, so the squares on a crowded channel will be smaller than the
squares on a channel with fewer stars.  We find that 
ten stars is sufficient to create
PRFs that capture the cores with good quality.

\begin{figure}
\epsscale{1}
\plotone{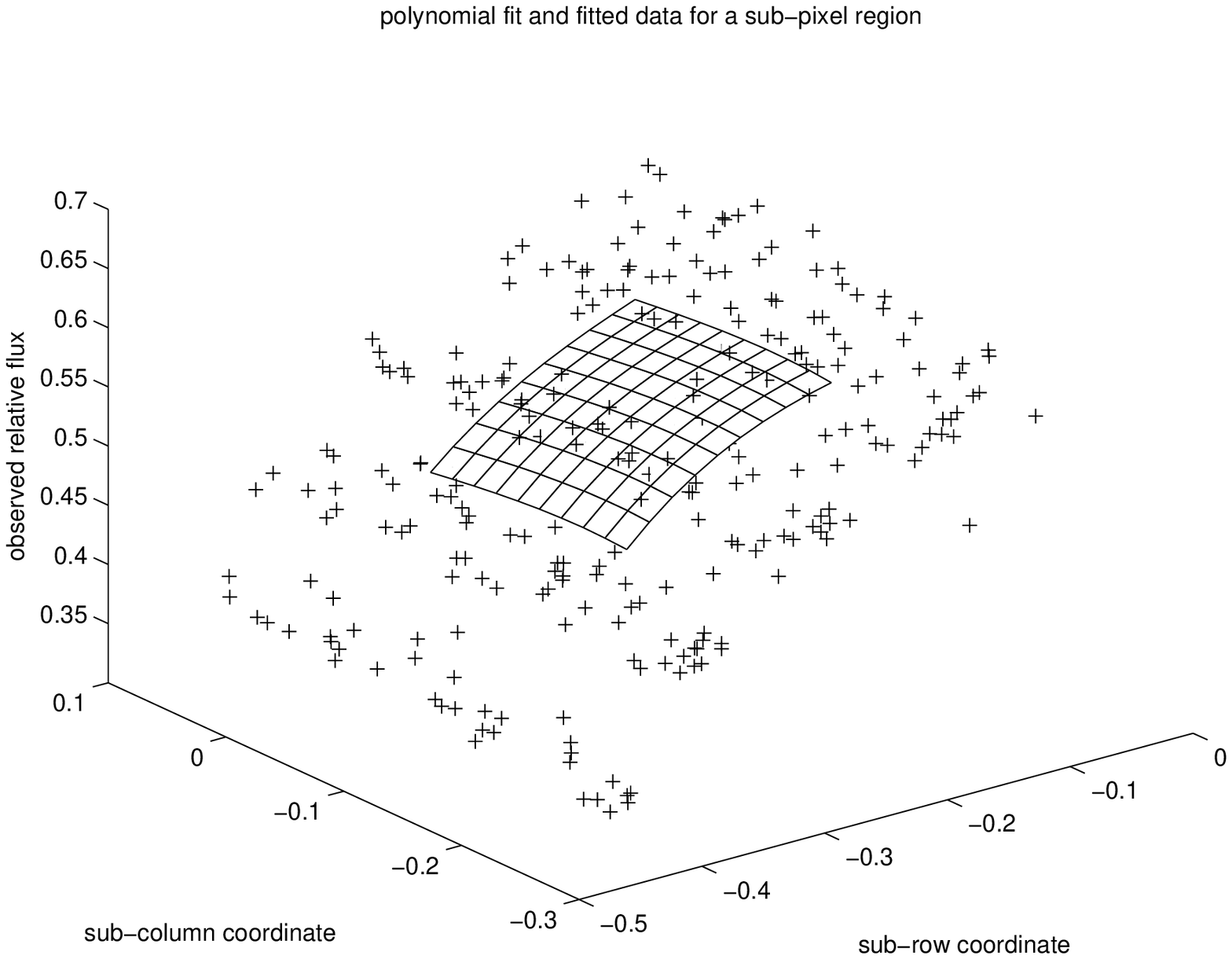}
\caption{An example of a PRF polynomial patch fitted to data.  The domain of the data
is larger than the domain of the patch to reduce the size of discontinuities between patches.}\label{fig4}
\end{figure}

Once the data for a particular PRF is selected, the pixel data is grouped in two ways:
(1) The integer pixel coordinates of each pixel are registered onto the $n \times n$ PRF 
pixel array so that the central pixel of the star corresponds to the central pixel of
the PRF pixel array.  (2) Each star's pixel is placed into a sub-pixel bin corresponding 
to the sub-pixel region containing that star.  The sub-pixel
bin covers a larger area than the sub-pixel region, providing overlapping data used for the 
polynomial fit, minimizing discontinuities between adjacent polynomial patches.
For each pixel in the PRF pixel array, and for each sub-pixel region in that pixel,
the $k$th pixel value $p_k$ is collected with its sub-pixel coordinates $\left(x_k,y_k\right)$.  
The PRF polynomial patch $\mathrm{PRF}\left(x_k,y_k\right)$ for that sub-pixel region is 
the polynomial that minimizes
\begin{equation}
\chi^{2}=\sum_{k=1}^{N}\left[ \frac{1}{\sigma_{k}}\left(
p_{k}-\mathrm{PRF}\left( x_{k},y_{k}\right)
\right) \right] ^{2}.  \label{eqn:prf_least_squares}
\end{equation}
Here $N$ is the number of pixel values in this sub-pixel region for this pixel, and 
$\sigma_{k}$ is the uncertainty in the measurement of $p_k$ (Figure~\ref{fig4}).  

The order of the polynomial is determined by
the modified Akaike information criterion \citep{akaike74}.  
If 
\begin{equation}
\mu\left(o\right)=\frac{1}{N}\sum_{k=1}^{N}\left(
p_{k}-\mathrm{PRF}_{o}\left( x_{k},y_{k}\right)
\right) ^{2}.  \label{eqn:prf_mean-square-error}
\end{equation}
is the mean square error of the polynomial fit $\mathrm{PRF}_{o}$ for a given order $o$, then Akaike's
modified information criterion selects the order that minimizes 
\begin{equation}
2c + N \log\left(\mu\left(o\right)\right) + \frac{2c\left(c-1\right)}{N-c-1}  \label{eqn:prf_aic}
\end{equation}
where $c$ is the number of coefficients in the polynomial $\mathrm{PRF}_{o}$ for order $o$.
The polynomial order will vary from sub-pixel region to sub-pixel region.  
The minimal solution to equation (\ref{eqn:prf_least_squares}) is found using a fitting method that
is robust against outliers.  

A difficulty was encountered when computing PRFs in a five-PRF set.  
While the data was sufficient to capture the core of the PRF, the far wings were highly 
vulnerable to background pollution from dim background stars due to the small number of targets used.
In contrast, the single PRFs computed on the same output
channel using all the targets had, for the most part, well-behaved wings (we occasionally hand-edited the single 
PRF far wings to remove obvious background pollution).  
The solution was to replace polynomials in the far wings of the PRFs in the 5-PRF set
with those from the corresponding pixel and sub-pixel regions from the single PRF.  The region of 
replacement was determined by computing a contour of a manually determined background value 
that encloses the PRF core in each PRF of the five-PRF set.
All polynomial patches outside that contour were replaced by polynomials from the single PRF.  
The resulting discontinuities 
were of the same magnitude as the discontinuities that already existed between polynomial patches.  

The resulting PRFs do a good job of simulating pixel flux from the stars in the KIC, as shown in 
Figure~\ref{fig2}.  We find that central pixel energies vary across the \emph{Kepler} focal plane, with 11\% 
of the output channels having central pixel energies less than 0.3, 20\% between 
0.3 and 0.4, 37\% between 
0.4 and 0.5, 28\% between 
0.5 and 0.6, and 4\% between 0.6 and 0.7.  

There are several possible sources of PRF error, including: 
\begin{itemize}
\item Changes in the statistics of spacecraft pointing jitter.  This has remained stationary over time
but is being monitored.
\item Focus has changed slightly but measurably over time due to water desorption and 
thermal variations \citep{jenkins10a}.
\item KIC errors, including variable stars and blends, can distort the measured PRF and cause errors for individual stars.
\item CCD non-uniformities.
\item The PRF computation did not attempt to include effects due to star color.  
Pre-flight simulated PRFs based on Code V simulations of the optical
PSF from Ball Aerospace and Technologies Corporation indicate that color can have a non-trivial effect 
on the PRF.  
\end{itemize}
The impact and mitigation of these errors is an area of active investigation.

%%%%%%%%%%%%%%%%%%%%%%%%%%%%%%%%%%%%
%%%%%%%%%%%%%%%%%%%%%%%%%%%%%%%%%%%%
\subsection{The FPG/PRF Iteration} \label{section:prfFpgIteration}

The computation described in \S\ref{section:prfComputation} assumes
that the pixel data have been calibrated, had their background removed, and that
that \emph{Kepler} attitude and FPG model are known.  Pixel data calibration 
and background removal are performed by the same software from the \emph{Kepler} SOC pipeline that
is used to process \emph{Kepler} science data \citep{jenkins10b}.  The \emph{Kepler} attitude and FPG model,
however, are based on PRF centroiding (\S\ref{section:stellarCentroids}).
Therefore an iterative approach is required.  

Two SOC modules are used: CAL, which performs pixel-level calibration, and
PA which performs background characterization and removal as well as 
stellar centroiding and computation of motion polynomials (PA also performs
aperture photometry during \emph{Kepler} science processing, but PRF does not
use this feature).  Background-removed pixels are not persisted in the SOC pipeline,
so the background removal function of PA must be called with each iteration.

PRF processing begins with a non-iterative bootstrap phase: 
Pixel data is calibrated by CAL, then PA is called with a pre-launch
estimate of the PRF and FPG models that are used for centroiding and 
motion polynomial creation. At this point the iterative loop is entered:  the FPG calculation 
provides an updated FPG model and attitude solution; PA performs background removal;
the PRF computation described in \S\ref{section:prfComputation} is performed; 
PA performs centroiding and motion polynomial computation using the updated PRF. The iteration 
is repeated until the FPG model is observed to converge. 

To obtain convergence, it was necessary to ``pin'' the centroid of the PRF 
computed in \S\ref{section:prfComputation} to a specific location because
both PRF and FPG can move the location of a star image on a CCD.  We chose
to constrain the PRF so that the value-weighted centroid of all input pixel values is at the
center of the PRF array.

%%%%%%%%%%%%%%%%%%%%%%%%%%%%%%%%%%%%
%%%%%%%%%%%%%%%%%%%%%%%%%%%%%%%%%%%%
%%%%%%%%%%%%%%%%%%%%%%%%%%%%%%%%%%%%
%%%%%%%%%%%%%%%%%%%%%%%%%%%%%%%%%%%%
%%%%%%%%%%%%%%%%%%%%%%%%%%%%%%%%%%%%
%%%%%%%%%%%%%%%%%%%%%%%%%%%%%%%%%%%%
\section{PRF Applications} \label{section:prfApplications}

%The \emph{Kepler} PRF can be used for a variety of purposes.  This section describes
%the two most important applications: the selection of optimal pixels for photometry
%and the determination of stellar centroid locations via PRF fitting.
%%%%%%%%%%%%%%%%%%%%%%%%%%%%%%%%%%%%
%%%%%%%%%%%%%%%%%%%%%%%%%%%%%%%%%%%%
\subsection{Optimal Pixel Selection} \label{section:optimalPixels}

There is a limit of $5.4 \times 10^6$ pixels per 
\emph{Kepler} long cadence.  Among these pixels are those chosen for stellar \emph{Kepler}
targets \citep{bathala10}.  These pixels are selected to support high-precision aperture photometry 
of these targets: for each stellar target those pixels are selected which maximize 
the signal-to-noise ratio (SNR) for that target.

The selection process relies on synthetic images generated using the PRF, KIC
and a zodiacal light model.  
For each target, two synthetic images are created: 1) an image that contains the target star only (``target'')
and 2) all stars except the target star, plus zodiacal light (``scene'').  Both images
include effects of saturation, charge transfer efficiency and smear induced by the lack of a shutter.  
The noise for each pixel $p$ (in electrons) contains contributions from shot, read ($\sigma_\mathrm{read}$) 
and quantization noise ($\sigma_\mathrm{quant}$) \citep{caldwell10} and is computed as 
$\sigma = \sqrt{p_\mathrm{target} + p_\mathrm{scene} + \sigma_\mathrm{read}^2 +  \sigma_\mathrm{quant}^2}$.
The SNR of the pixel is $p_\mathrm{target}/\sigma$.

The pixels are sorted in decreasing order of SNR.  Starting with the first pixel, as each 
pixel is added the SNR of the collection of pixels is computed, and
the collection of pixels that maximize the collection's SNR defines the optimal aperture for this target.
Margin against PRF and KIC errors is provided by a ring of pixels placed around the optimal aperture, as well
as a column of pixels on the left for undershoot correction \citep{haas10}.
The result is the \emph{requested aperture} for storage and downlinking. 

\begin{figure}
\epsscale{1}
\plotone{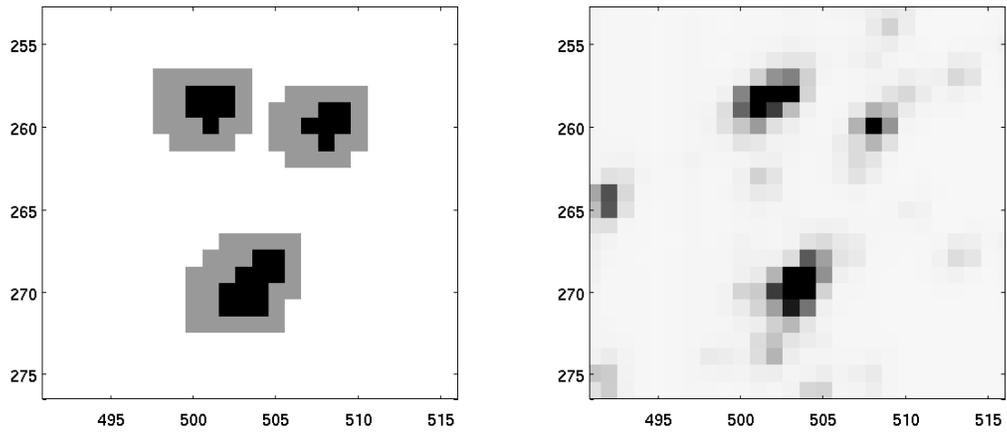}
\caption{Optimal and spacecraft apertures for three \emph{Kepler} targets.  Left: the optimal apertures for 
the targets in black and the spacecraft apertures in grey.  Right: the synthetic image containing 
the targets and other stars.}\label{fig5}
\end{figure}

Software on the \emph{Kepler} spacecraft supports 1024 arbitrarily shaped \emph{spacecraft apertures}, far fewer than
the number of requested apertures.  Therefore each requested aperture is assigned to the spacecraft aperture
with the smallest number of pixels that contains the requested aperture.  
Figure~\ref{fig5} shows example optimal and spacecraft apertures.

A synthetic image containing all stars in the KIC is used to determine background pixels that are 
stored, downlinked and used to build and remove a background model in SOC processing
of the science targets.

%%%%%%%%%%%%%%%%%%%%%%%%%%%%%%%%%%%%
%%%%%%%%%%%%%%%%%%%%%%%%%%%%%%%%%%%%
\subsection{Stellar Centroids} \label{section:stellarCentroids}

Centroid locations for unsaturated, sufficiently bright target stars are determined by
PRF fitting: a non-linear fit is performed which 
varies the location and flux of the PRF, minimizing the $\chi^2$ of the pixel-wise difference 
of the PRF-based pixel image (\S\ref{section:prfToPixels}) 
and the flight pixel values for that target.  The resulting centroids show 
high stability, with changes over time of about $0.1$ millipixels 
(after removing systematic effects).  Errors in 
the PRF model induce local biases, which exhibit a median of about 1 millipixel and 
median absolute deviation
of ~22 millipixels pixels, across the focal plane.

%%%%%%%%%%%%%%%%%%%%%%%%%%%%%%%%%%%%
%%%%%%%%%%%%%%%%%%%%%%%%%%%%%%%%%%%%
\acknowledgments

We thank the larger \emph{Kepler} team for their support and hard work.  
Funding for the \emph{Kepler} mission is provided by NASA, Science Mission Directorate.

{\it Facilities:} \facility{Kepler}.




\clearpage

%% the file, uncomment the \plotone, \plottwo, and
%% If you need a layout that cannot be achieved with \plotone or
%% \plottwo, you can invoke the graphicx package directly with the
%% \includegraphics command or use \plotfiddle. For more information,
%% please see the tutorial on "Using Electronic Art with AASTeX" in the
%% documentation section at the AASTeX Web site,
%% http://www.journals.uchicago.edu/AAS/AASTeX.
%%
%% The examples below also include sample markup for submission of
%% supplemental electronic materials. As always, be sure to check
%% the instructions to authors for the journal you are submitting to
%% for specific submissions guidelines as they vary from

%% This example uses \plotone to include an EPS file scaled to
%% 80% of its natural size with \epsscale. Its caption
%% has been written to indicate that additional figure parts will be
%% available in the electronic journal.

%% The following command ends your manuscript. LaTeX will ignore any text

\end{document}